# Determination of protein structural ensembles using cryo-electron microscopy

Massimiliano Bonomi[*, ‡] and Michele Vendruscolo[*]

*Department of Chemistry, University of Cambridge, Cambridge CB2 1EW, United Kingdom*

*To whom correspondence should be addressed: massimiliano.bonomi@pasteur.fr; mv245@cam.ac.uk

‡Current address: Structural Bioinformatics Unit, Institut Pasteur, CNRS UMR 3528, 75015 Paris, France

**Abstract**

Achieving a comprehensive understanding of the behaviour of proteins is greatly facilitated by the knowledge of their structures, thermodynamics and dynamics. All this information can be provided in an effective manner in terms of structural ensembles. A structural ensemble can be obtained by determining the structures, populations and interconversion rates for all the main states that a protein can occupy. To achieve this goal, integrative methods that combine experimental and computational approaches provide powerful tools. Here we focus on cryo-electron microscopy, which has become over recent years an invaluable resource to bridge the gap from order to disorder in structural biology. In this review, we provide a perspective of the current challenges and opportunities in determining protein structural ensembles using integrative approaches that can combine cryo-electron microscopy data with other available sources of information, along with an overview of the tools available to the community.

**Understanding protein behaviour**

All proteins are to some extent dynamic as they populate a variety of different states [1,2]. Such behaviour is often required to achieve specific functions [3] and encompasses a wide range of time and length scales, from side-chains motions [4] and loop jiggling [5] to larger conformational changes [6]. In many cases, proteins present entirely disordered regions that can extend to encompass their entire sequences, as in the case of disorder proteins [7]. Proteins in this class constitute perhaps one-third of the human proteome, are highly involved in regulation and signaling processes [8], and play a central role in neurodegenerative disorders [9]. More generally, every biological system exhibits a spectrum of conformational motions that spans a continuum between order and disorder [10].

As a consequence, understanding the behaviour of proteins requires the simultaneous determination of the structures of all the relevant states that these molecules can occupy under certain external conditions, their populations (i.e., their *thermodynamics*), and the rates of interconversion among these states (i.e., their *dynamics*). Therefore, the problem that one should strive to solve to shed light into protein behaviour is the determination of a structural ensemble (**Figure 1A**), as defined by a set of structures (**Figure 1B**), along with the statistical weights quantifying their relative populations (**Figure 1C**) and the transition rates measuring dynamics (**Figure 1D**). The central question thus becomes: operatively, how can one determine these structural ensembles?

**Integrative methods for protein structural ensemble determination**

In recent years, integrative (or hybrid) methods that combine computational and experimental approaches have been proven very successful in determining the structures of complex biological systems at various resolutions [11,12]. These methods have also been extended to enable the determination of structural ensembles using ensemble-averaged experimental data, such as those provided by nuclear magnetic resonance (NMR) spectroscopy [13] and small angle X-ray scattering (SAXS) [14], or distributions of experimental observations over the ensemble, such as Förster resonance energy transfer (FRET) [15] or double electron-electron resonance (DEER) data [16]. Without entering into the details of these methods, which are covered in many excellent reviews [17-22], integrative ensemble approaches are aimed at properly combining all the information available to overcome the limitations of individual techniques and maximize the accuracy of the resulting structures or structural ensembles. In most cases, a physico-chemical *a priori* description of the system (e.g. a force field) is augmented by additional conformational restraints aimed at enforcing an agreement of the structural ensemble with the available experimental data [17]. These terms are encoded into a hybrid energy function, which is used to drive sampling methods like Monte Carlo (MC) or molecular dynamics (MD) and ultimately to generate structural models consistent with the input information [17].

**Challenges in the implementation of protein structural ensemble determination methods**

The procedure outlined above presents three fateful challenges (**Figure 2**), whose importance should not be underestimated [18]. The first one is obtaining an accurate hybrid energy function (**Figure 2A**). Current force fields for MC and MD simulations, despite continuous progress [23,24], still provide only an approximate *prior* description of the system, and therefore they should be complemented by highly-informative experimental data. The second challenge is that experimental data, as well as the theoretical models used to predict the data from a conformation and assess the consistency with experimental observations (i.e., the *predictors* or *forward models*), are always affected by random and systematic errors [25]. Quantifying such errors is crucial to avoid enforcing a too strong agreement with noisy or incorrect data, and ultimately to properly integrate all pieces of information together into a well-balanced energy function (**Figure 2B**). The third challenge is that the free energy landscape of a complex system is almost invariably extremely difficult to sample at reasonable computational cost by MC or MD methods, as these simulations can easily remain trapped in local free energy minima. Therefore, the use of enhanced-sampling techniques [26], such as umbrella sampling [27], metadynamics [28] or replica-exchange [29], is almost always necessary to exhaustively explore the conformational landscape (**Figure 2C**).

Provided that all these challenges are successfully overcome, integrative methods provide excellent tools to maximize the use of all the information available, to model accurate structural ensembles, and ultimately to obtain exquisite insights into protein behaviour.

**The cryo-electron microscopy revolution**

Among the different types of experimental data that can be used in integrative modelling, cryo-electron microscopy (cryo-EM) has become an invaluable source of information [30]. Stimulated by the recent developments in both instrumentation and image-processing software [31,32], over the last years cryo-EM has allowed the structures of complexes and individual proteins of extraordinary biological importance to be determined at nearly-atomic resolution, thus establishing itself as a powerful technique in structural biology [33,34] .

Cryo-EM has the potential to provide an exhaustive characterization of the entire conformational landscape of complex macromolecular systems. In principle, this result can be directly obtained from the raw data, *i.e.* the set of single-particle, two-dimensional (2D) images of the system, which is deposited on a thin layer of vitreous ice and imaged in different orientations and conformations (**Figure 3A**). Unfortunately, these single-particle images usually have a low signal-to-noise ratio and therefore additional post-processing steps are typically performed to obtain higher resolution 2D images (class-averages, **Figure 3B**) and ultimately one or more three-dimensional (3D) reconstructions (density maps, **Figure 3C**) [35,36].

The image-processing of the raw cryo-EM data has the desirable outcome of generating a small number of density maps, which represent the system in distinct conformational states [37,38], often at high resolution. These density maps provide a picture of the highly-populated free energy minima in the conformational landscape, but intermediates along the pathways that connect these states might remain elusive. In this regard, integrating cryo-EM data with molecular simulations or other experimental data might help connecting the dots. In addition, in many cases part of the conformational space visible in the 2D single-particle images is discarded, and thus lost, or averaged out in the classification and reconstruction process. In this situation, low-resolution areas of high-resolution density maps might conceal highly dynamic parts of the system, provided that these regions result from the averaging of images in different conformations and not from the noise caused, for example, by radiation damage. When focused classification [39,40] of the highly dynamic regions is not able to resolve individual conformations, integrative ensemble approaches [17-22] can be particularly helpful for providing an exhaustive characterization of the conformational landscape of the system.

**Determination of protein structural ensembles with cryo-EM data**

We will not focus our attention on maximum-likelihood and Bayesian methods for 2D particle classification or 3D reconstruction in conformationally heterogenous systems, nor on approaches that are aimed at determining a single structural model from 2D images or 3D reconstructions, as these methods are already covered in other recent reviews [36,41]. We note in particular that in addition to individual structures, many of these methods [42-54] generate ensembles of structural models that reflect the limited information available on the systems, and thus the fact that different models might be equally consistent with the input data. We refer to these ensembles as *uncertainty ensembles* [17]. The characteristic of these ensembles is that they converge more and more to individual structures as the amount of experimental information used to derive them is increased. These uncertainty ensembles, however, are not aimed to reflect the conformational heterogeneity arising from the internal dynamics of the systems, which is instead described by *thermodynamic ensembles*. These ensembles are described in **Figure 1**. This section will instead provide an overview of the approaches that model thermodynamic ensembles using cryo-EM 2D and 3D data along with other available information and that can deal with systems with a continuous spectrum of dynamics. Particular attention will be given to those methods that are implemented and distributed in open-source, freely-available software.

BioEM [55] is a Bayesian approach that uses a probabilistic framework to assess the consistency between a structural model and a set of single-particle cryo-EM images. The major advantages of this approach are the use of the raw data (**Figure 3A**), before any clustering or averaging procedure, and the fact that it accounts for uncertainty in the position and orientation of the system as well as noise in the experimental images. A maximum-parsimony approach [17] can also be used to identify a minimal ensemble that can collectively explain the data, as demonstrated in the case of the ESCRT I–II supercomplex [55]. The method is implemented in a highly-parallelized and GPU-accelerated open source software [56], freely-available at https://bioem.readthedocs.io.

The EMageFit approach [57] is based on a maximum-likelihood score that quantifies the agreement with 2D class averages (**Figure 3B**), measured in terms of the cross-correlation between experimental and predicted images, as well as with other experimental data and any prior information available. This method has been used to determine the conformational states of the human TfR-Tf complex [57], in combination with cross-linking/mass-spectrometry data and proximity information, and of the architecture of the yeast spindle pole body [58], in combination with *in vivo* FRET, yeast two-hybrids, SAXS, and X-ray crystallography data. EMageFit is implemented in the Integrative Modeling Platform (IMP, http://integrativemodeling.org) [59], an open-source, freely-available library for integrative modelling using a wide variety of experimental data.

The Mosaics–EM method [60] utilizes a MC sampling procedure with natural moves, combined with a simulated annealing protocol [61], to refine conformations using 2D class averages (**Figure 3B**). Similarly to EMageFit, this approach relies on a hybrid-energy function to quantify the agreement of a model with respect to the experimental 2D class averages in terms of cross-correlation between target and model projections. The method was capable of resolving the conformational heterogeneity of the *Methonococcus maripaludis* chaperonin system [60], which is known to populate a mixture of open and close states.

The recently proposed metainference approach [62] combines a general Bayesian method to model structural ensembles using noisy data with a maximum entropy approach to take into account the ensemble-averaged nature of experimental data [13,63]. While metainference was initially applied to the characterization of disordered systems using NMR data [64], it has recently been extended to model conformational ensembles using cryo-EM 3D maps (**Figure 3C**) [65,66]. The goal of this method is to unravel the conformational heterogeneity hidden in low-resolution areas of high-resolution maps and to properly combine cryo-EM with other experimental data. Therefore, it is particularly useful for characterizing highly flexible regions of a system, when focused classification methods might fail. Metainference for cryo-EM (EMMI) is based on a Bayesian probabilistic framework that accounts for both ensemble-averaging over multiple conformations and the presence of non-uniform noise across the map. Sampling is carried out by MD, possibly in combination with metadynamics [28] to accelerate the exploration of the conformational space [67]. EMMI has been applied to the simultaneous determination of structure and dynamics of the STRA6 integral membrane receptor [66] and of the ClpP protease [68]. The approach is implemented in the ISDB module [69] of the open-source PLUMED library [70], freely available at www.plumed.org.

**Integrating cryo-EM with other methods to improve the quality of protein structural ensembles**

Integrating medium- and low-resolution cryo-EM data (and similarly X-ray crystallography data in the 3-5 Å resolution range [71]) with other sources of information enables one to overcome the issue of data sparseness [25] and thus to obtain more accurate single-structure models. Some examples include the combination of cryo-EM with NMR [72], SAXS [73], and cross-linking/mass-spectrometry [74] data. In this regard, Bayesian inference [65,75,76] provides a robust framework to objectively weight each piece of information used in the modelling by automatically quantifying its level of noise or uncertainty.

As the resolution of cryo-EM data is rapidly approaching atomistic detail, one could ask in which situations we can still benefit from integrating cryo-EM with other sources of information. One case is when the single-particle images, which typically have a low signal-to-noise ratio, are used, as for example in the BioEM approach [55]. Another example is when high-resolution 3D maps present areas at lower resolution, due to averaging over multiple conformations during the classification and reconstruction process, as used in the EMMI approach [65,66]. In both cases, integrating cryo-EM with other experimental data or more accurate prior information (such as explicit descriptions of solvent and membrane environments) can help increase the accuracy of the reconstructed ensemble. One particularly intriguing combination is with NMR observables, which in most cases, such as chemical shifts and residual dipole couplings, report average quantities over an ensemble of conformations.

These ensemble-averaged data can be optimally integrated with cryo-EM in the metainference approach [62] or using other reweighting techniques [77] to modify the ensemble obtained from cryo-EM data alone [55]. This integration should be carried out with particular care, as NMR averaging depends on the time scale of interconversion among different states. Therefore, it is essential to estimate kinetic properties on the cryo-EM-derived structural ensemble prior to properly incorporate ensemble-averaged NMR data.

A representative example of the opportunities offered by integrative approaches is the recent characterization of the structure and dynamics of the ClpP protease by a combined cryo-EM and NMR strategy [68]. In this study, the EMMI approach has been used to generate a conformational ensemble (**Figure 3D**) using cryo-EM data at 3.6 Å resolution (**Figure 3C**). This ensemble highlighted the presence of significant dynamics in the N-terminal gating region of the protease (**Figure 3E**), which was resolved at lower resolution by cryo-EM (**Figure 3C**). NMR methyl chemical shifts and paramagnetic relaxation enhancement data were back-calculated from the cryo-EM ensemble and compared to the experimental measurements, showing a good agreement. This example can be regarded as a first step toward the integration of cryo-EM with NMR data to generate accurate conformational ensembles that are revealing of the dynamics, and ultimately function, of a complex biological system.

Finally, MD simulations can complement the characterization of the conformational landscape of a system when multiple cryo-EM maps are first used to model individual structures representing local free-energy minima. By connecting these landmarks using standard or enhanced-sampling MD simulations, one can obtain a more complete overview of the conformational landscape of complex biological systems [78].

**Community challenges in the validation and dissemination of protein structural ensembles**

The recent developments in the cryo-EM field, along with the new advances in integrative modelling approaches for protein structural ensemble determination, are exposing new challenges that the community should collectively strive to solve.

1) Researchers should be able to readily access data and information about specific experiments, such as the raw single-particle images and details about the post-processing of the raw data into 3D density maps. This information might be required to use, or further develop, specific integrative approaches [55,56]. An important initial step in this direction is the Electron Microscopy Public Image Archive (EMPIAR) for raw, 2D electron microscopy images (https://www.ebi.ac.uk/pdbe/emdb/empiar/) [79].

2) Modelling procedures and protocols for structural ensemble determination should be shared within the community in order to encourage reproducibility, which can be further facilitated by using well-established approaches for structural ensemble comparison [80]. Furthermore, the resulting structural ensembles should be easily accessible. In this regard, a pioneering effort is the creation of the PDB-Dev prototype repository for integrative structural models (https://pdb-dev.wwpdb.org/) [81], which hosts multi-scale, multi-state and time-ordered ensembles of macromolecular assemblies along with the data used in their modelling. This information should include structural models, possibly after clustering integrative simulations into a set of discrete states using community-established criteria, states populations and transition rates (**Figure 1**).

3) Effective structural ensemble validation procedures should be established and progressively enforced when depositing structural ensembles to public repositories. These procedures should assess both the protocols used to generate the structural ensembles and their overall quality. First, exhaustive sampling of the conformational landscape, and therefore convergence of integrative simulations, should be monitored by either reporting the statistical error in the estimated state populations or more sophisticated approaches [82]. Second, structural ensembles should be evaluated in terms of their overall physico-chemical quality, as typically quantified by the Molprobity score [83] calculated by the PHENIX [84] validation suite, as well as the fit of the cryo-EM data, often measured in terms of cross-correlation between experimental and model maps or using the EMRinger [85] approach to assess the precise fitting of an atomic model into a cryo-EM map. These validation tools and the corresponding criteria used to establish when single-structure models are acceptable for deposition should therefore be adapted to deal with protein ensembles.

Additional validations of cryo-EM and integrative structural ensembles can be obtained by assessing the consistency between predictions calculated from the ensembles and independent experimental data not used in the modelling, such as NMR and SAXS data. Such comparison must be done with caution and upon considering: *i)* the kinetics of interconversion among different structural states and thus how experimental observables are averaged over the ensemble; *ii)* the errors in the predictors of experimental observables as well as in the experimental data; *iii)* the statistical uncertainty in the population of the states determined by integrative simulations; *iv)* the fact that cryo-EM and NMR/SAXS data are collected in different conditions, e.g. vitreous ice and solution at room temperature, therefore some differences between the underlying ensembles can be expected.

More extensive discussions of these challenges can be found in a recent Special Issue of the Journal of Structural Biology (http://challenges.emdataresource.org/?q=node/84).

**Conclusions**

Cryo-EM has become in recent years increasingly successful in structural biology because of its ability to provide accurate characterizations of the conformational landscapes of complex biological systems.

Alongside these developments, we have assisted to the flourishing of integrative modelling approaches to integrate cryo-EM with other sources of experimental and prior information [17-22]. These methods further increase the accuracy of the structural ensembles generated using cryo-EM data. As we have emphasised here, however, this goal can only be achieved if the challenges related to the accuracy of the hybrid energy function used, the treatment of all sources of errors and uncertainty, and the sampling issues of MD and MC simulations are successfully addressed (**Figure 2**).

Over the next years, we will assist to new technical developments in the cryo-EM field as well as in the data analysis and integrative modelling techniques, which will facilitate the synergistic use of all the available information to characterize structure and dynamics, and ultimately function, of ever more complex and important biological systems. Therefore, it is now the right time for the entire community to discuss how the validation methods and dissemination protocols developed over the years for single-structure models should be revisited and updated for the era of protein structural ensembles.

**Acknowledgements**

We are grateful to Alessandro Barducci and Francesco Aprile for their feedback on Figure 1, to Giulia Vecchi for preparing Figure 2, to Zev Ripstein for providing the ClpP cryo-EM data visualized in panels A, B, and C of Figure 3, and to James Fraser for carefully reading the manuscript.

**References**

1. Henzler-Wildman K, Kern D: **Dynamic personalities of proteins**. *Nature* 2007, **450**:964-972.

2. Mittermaier A, Kay LE: **New tools provide new insights in NMR studies of protein dynamics**. *Science* 2006, **312**:224-228.

3. Frauenfelder H, McMahon B: **Dynamics and function of proteins: The search for general concepts**. *Proc Natl Acad Sci USA* 1998, **95**:4795-4797.

4. Trbovic N, Cho J-H, Abel R, Friesner RA, Rance M, Palmer III AG: **Protein Side-Chain Dynamics and Residual Conformational Entropy**. *J Am Chem Soc* 2009, **131**:615-622.

5. Taylor SS, Kornev AP: **Protein Kinases: Evolution of Dynamic Regulatory Proteins**. *Trends Biochem Sci* 2011, **36**:65-77.

6. Krukenberg KA, Street TO, Lavery LA, Agard DA: **Conformational dynamics of the molecular chaperone Hsp90**. *Q Rev Biophys* 2016, **44**:229-255.

7. Tompa P: **Intrinsically disordered proteins: a 10-year recap**. *Trends Biochem Sci* 2012, **37**:509-516.

8. Wright PE, Dyson HJ: **Intrinsically disordered proteins in cellular signalling and regulation**. *Nat Rev Mol Cell Biol* 2015, **16**:18-29.

9. Uversky VN, Oldfield CJ, Dunker AK: **Intrinsically disordered proteins in human diseases: introducing the D2 concept**. *Annu Rev Biophys* 2008, **37**:215-246.

10. Sormanni P, Piovesan D, Heller GT, Bonomi M, Kukic P, Camilloni C, Fuxreiter M, Dosztanyi Z, Pappu RV, Babu MM, et al.: **Simultaneous quantification of protein order and disorder**. *Nat Chem Biol* 2017, **13**:339-342.

11. van den Bedem H, Fraser JS: **Integrative, dynamic structural biology at atomic resolution--it's about time**. *Nat Methods* 2015, **12**:307-318.

12. Ward AB, Sali A, Wilson IA: **Biochemistry. Integrative structural biology**. *Science* 2013, **339**:913-915.

13. Lindorff-Larsen K, Best RB, Depristo MA, Dobson CM, Vendruscolo M: **Simultaneous determination of protein structure and dynamics**. *Nature* 2005, **433**:128-132.

14. Bernadó P, Mylonas E, Petoukhov MV, Blackledge M, Svergun DI: **Structural characterization of flexible proteins using small-angle X-ray scattering**. *J Am Chem Soc* **129**:5656-5664.

15. Dimura M, Peulen TO, Hanke CA, Prakash A, Gohlke H, Seidel CAM: **Quantitative FRET studies and integrative modeling unravel the structure and dynamics of biomolecular systems**. *Curr Opin Struct Biol* 2016, **40**:163-185.

16. Roux B, Islam SM: **Restrained-Ensemble Molecular Dynamics Simulations Based on Distance Histograms from Double Electron- Electron Resonance Spectroscopy**. *J Phys Chem B* 2013, **117**:4733-4739.

17. Bonomi M, Heller GT, Camilloni C, Vendruscolo M: **Principles of protein structural ensemble determination**. *Curr Opin Struct Biol* 2017, **42**:106-116.

18. Bottaro S, Lindorff-Larsen K: **Biophysical experiments and biomolecular simulations: A perfect match?** *Science* 2018, **361**:355-360.

19. Fisher CK, Stultz CM: **Constructing ensembles for intrinsically disordered proteins**. *Curr Opin Struct Biol* 2011, **21**:426-431.

20. Ravera E, Sgheri L, Parigi G, Luchinat C: **A critical assessment of methods to recover information from averaged data**. *Phys Chem Chem Phys* 2016, **18**:5686-5701.

21. Cesari A, Reißer S, Bussi G: **Using the maximum entropy principle to combine simulations and solution experiments**. *Computation* 2018, **6**:15.

22. Boomsma W, Ferkinghoff-Borg J, Lindorff-Larsen K: **Combining Experiments and Simulations Using the Maximum Entropy Principle**. *Plos Comp Biol* 2014, **10**.

23. Nerenberg PS, Head-Gordon T: **New developments in force fields for biomolecular simulations**. *Curr Opin Struct Biol* 2018, **49**:129-138.

24. Robustelli P, Piana S, Shaw DE: **Developing a molecular dynamics force field for both folded and disordered protein states**. *Proc Natl Acad Sci USA* 2018: 10.1073/pnas.1800690115.

25. Schneidman-Duhovny D, Pellarin R, Sali A: **Uncertainty in integrative structural modeling**. *Curr Opin Struct Biol* 2014, **28**:96-104.

26. Abrams C, Bussi G: **Enhanced Sampling in Molecular Dynamics Using Metadynamics, Replica-Exchange, and Temperature-Acceleration**. *Entropy* 2014, **16**:163-199.

27. Torrie GM, Valleau JP: **Non-Physical Sampling Distributions in Monte-Carlo Free-Energy Estimation - Umbrella Sampling**. *J Comp Phys* 1977, **23**:187-199.

28. Laio A, Parrinello M: **Escaping free-energy minima**. *Proc Natl Acad Sci USA* 2002, **99**:12562-12566.

29. Sugita Y, Okamoto Y: **Replica-exchange molecular dynamics method for protein folding**. *Chem Phys Lett* 1999, **314**:141-151.

30. Kuhlbrandt W: **The resolution revolution**. *Science* 2014, **343**:1443-1444.

31. Bai XC, McMullan G, Scheres SH: **How cryo-EM is revolutionizing structural biology**. *Trends Biochem Sci* 2015, **40**:49-57.

32. Glaeser RM: **How good can cryo-EM become?** *Nat Methods* 2016, **13**:28-32.

33. Callaway E: **The Revolution Will Not Be Crystallized**. *Nature* 2015, **525**:172-174.

34. Nogales E: **The development of cryo-EM into a mainstream structural biology technique**. *Nat Methods* 2016, **13**:24-27.

35. Sigworth FJ: **Principles of cryo-EM single-particle image processing**. *Microscopy* 2016, **65**:57-67.

36. Cossio P, Hummer G: **Likelihood-based structural analysis of electron microscopy images**. *Curr Opin Struct Biol* 2018, **49**:162-168.

37. Zhao J, Benlekbir S, Rubinstein JL: **Electron cryomicroscopy observation of rotational states in a eukaryotic V-ATPase**. *Nature* 2015, **521**:241-245.

38. Cianfrocco MA, Kassavetis GA, Grob P, Fang J, Juven-Gershon T, Kadonaga JT, Nogales E: **Human TFIID binds to core promoter DNA in a reorganized structural state**. *Cell* 2013, **152**:120-131.

39. Scheres SH: **RELION: implementation of a Bayesian approach to cryo-EM structure determination**. *J Struct Biol* 2012, **180**:519-530.

40. Scheres SH: **Processing of structurally heterogeneous cryo-EM data in RELION**. *Methods Enzymol* 2016, **579**:125-157.

41. Lopez-Blanco JR, Chacon P: **Structural modeling from electron microscopy data**. *Wiley Interdiscip Rev Comput Mol Sci* 2015, **5**:62-81.

42. Rossmann MG, Bernal R, Pletnev SV: **Combining electron microscopic with X-ray crystallographic structures**. *Journal of Structural Biology* 2001, **136**:190-200.

43. Lasker K, Topf M, Sali A, Wolfson HJ: **Inferential Optimization for Simultaneous Fitting of Multiple Components into a CryoEM Map of Their Assembly**. *Journal of Molecular Biology* 2009, **388**:180-194.

44. Zheng WS: **Accurate Flexible Fitting of High-Resolution Protein Structures into Cryo-Electron Microscopy Maps Using Coarse-Grained Pseudo-Energy Minimization**. *Biophysical Journal* 2011, **100**:478-488.

45. Trabuco LG, Villa E, Mitra K, Frank J, Schulten K: **Flexible fitting of atomic structures into electron microscopy maps using molecular dynamics**. *Structure* 2008, **16**:673-683.

46. Topf M, Lasker K, Webb B, Wolfson H, Chiu W, Sali A: **Protein structure fitting and refinement guided by cryo-EM density**. *Structure* 2008, **16**:295-307.

47. Pandurangan AP, Vasishtan D, Alber F, Topf M: **gamma-TEMPy: Simultaneous Fitting of Components in 3D-EM Maps of Their Assembly Using a Genetic Algorithm**. *Structure* 2015, **23**:2365-2376.

48. Volkmann N, Hanein D: **Quantitative fitting of atomic models into observed densities derived by electron microscopy**. *Journal of Structural Biology* 1999, **125**:176-184.

49. Ratje AH, Loerke J, Mikolajka A, Brunner M, Hildebrand PW, Starosta AL, Donhofer A, Connell SR, Fucini P, Mielke T, et al.: **Head swivel on the ribosome facilitates translocation by means of intra-subunit tRNA hybrid sites**. *Nature* 2010, **468**:713-716.

50. Saha M, Morais MC: **FOLD-EM: automated fold recognition in medium- and low-resolution (4-15 A) electron density maps**. *Bioinformatics* 2012, **28**:3265-3273.

51. DiMaio F, Tyka MD, Baker ML, Chiu W, Baker D: **Refinement of Protein Structures into Low-Resolution Density Maps Using Rosetta**. *Journal of Molecular Biology* 2009, **392**:181-190.

52. Lindert S, Alexander N, Wotzel N, Karakas M, Stewart PL, Meiler J: **EM-Fold: De Novo Atomic-Detail Protein Structure Determination from Medium-Resolution Density Maps**. *Structure* 2012, **20**:464-478.

53. Habeck M: **Bayesian Modeling of Biomolecular Assemblies with Cryo-EM Maps**. *Front Mol Biosci* 2017, **4**:1-13.

54. Singharoy A, Teo I, McGreevy R, Stone JE, Zhao J, Schulten K: **Molecular dynamics-based refinement and validation for sub-5 A cryo-electron microscopy maps**. *Elife* 2016, **5**.

55. Cossio P, Hummer G: **Bayesian analysis of individual electron microscopy images: towards structures of dynamic and heterogeneous biomolecular assemblies**. *J Struct Biol* 2013, **184**:427-437.

56. Cossio P, Rohr D, Baruffa F, Rampp M, Lindenstruth V, Hummer G: **BioEM: GPU-accelerated computing of Bayesian inference of electron microscopy images**. *Comp Phys Comm* 2017, **210**:163-171.

57. Velazquez-Muriel J, Lasker K, Russel D, Phillips J, Webb BM, Schneidman-Duhovny D, Sali A: **Assembly of macromolecular complexes by satisfaction of spatial restraints from electron microscopy images**. *Proc Natl Acad Sci USA* 2012, **109**:18821-18826.

58. Viswanath S, Bonomi M, Kim SJ, Klenchin VA, Taylor KC, Yabut KC, Umbreit NT, Van Epps HA, Meehl J, Jones MH, et al.: **The molecular architecture of the yeast spindle pole body core determined by Bayesian integrative modeling**. *Mol Biol Cell* 2017, **28**:3298-3314.

59. Russel D, Lasker K, Webb B, Velazquez-Muriel J, Tjioe E, Schneidman-Duhovny D, Peterson B, Sali A: **Putting the pieces together: integrative modeling platform software for structure determination of macromolecular assemblies**. *PLoS Biol* 2012, **10**:e1001244.

60. Zhang J, Minary P, Levitt M: **Multiscale natural moves refinemacromolecules using single-particle electron microscopy projection images**. *Proc Natl Acad Sci USA* 2012, **109**:9845-9850.

61. Kirkpatrick S, Gelatt CD, Vecchi MP: **Optimization by Simulated Annealing**. *Science* 1983, **220**:671-680.

62. Bonomi M, Camilloni C, Cavalli A, Vendruscolo M: **Metainference: A Bayesian inference method for heterogeneous systems**. *Sci Adv* 2016, **2**:e1501177.

63. Cavalli A, Camilloni C, Vendruscolo M: **Molecular dynamics simulations with replica-averaged structural restraints generate structural ensembles according to the maximum entropy principle**. *J Chem Phys* 2013, **138**:094112.

64. Heller GT, Aprile FA, Bonomi M, Camilloni C, De Simone A, Vendruscolo M: **Sequence Specificity in the Entropy-Driven Binding of a Small Molecule and a Disordered Peptide**. *J Mol Biol* 2017, **429**:2772-2779.

65. Bonomi M, Hanot S, Greenberg C, Sali A, Nilges M, Vendruscolo M, Pellarin R: **Bayesian weighing of electron cryo-microscopy data for integrative structural modeling**. *Structure* 2018, **In press**.

66. Bonomi M, Pellarin R, Vendruscolo M: **Simultaneous Determination of Protein Structure and Dynamics Using Cryo-Electron Microscopy**. *Biophys J* 2018, **114**:1604-1613.

67. Bonomi M, Camilloni C, Vendruscolo M: **Metadynamic metainference: Enhanced sampling of the metainference ensemble using metadynamics**. *Sci Rep* 2016, **6**:31232.

68. Vahidi S, Ripstein ZA, Bonomi M, Yuwen T, Mabanglo MF, Juravsky JB, Rizzolo K, Velyvis A, Houry WA, Vendruscolo M, et al.: **Reversible inhibition of the ClpP protease via an N-terminal conformational switch**. *Proc Natl Acad Sci USA* 2018, **115**:E6447-E6456.

69. Bonomi M, Camilloni C: **Integrative structural and dynamical biology with PLUMED-ISDB**. *Bioinformatics* 2017, **33**:3999-4000.

70. Tribello GA, Bonomi M, Branduardi D, Camilloni C, Bussi G: **PLUMED 2: New feathers for an old bird**. *Comp Phys Comm* 2014, **185**:604-613.

71. McGreevy R, Singharoy A, Li Q, Zhang J, Xu D, Perozo E, Schulten K: **xMDFF: molecular dynamics flexible fitting of low-resolution X-ray structures**. *Acta Crystallogr D Biol Crystallogr* 2014, **70**:2344-2355.

72. Cuniasse P, Tavares P, Orlova EV, Zinn-Justin S: **Structures of biomolecular complexes by combination of NMR and cryoEM methods**. *Curr Opin Struct Biol* 2017, **43**:104-113.

73. Xu X, Yan C, Wohlueter R, Ivanov I: **Integrative Modeling of Macromolecular Assemblies from Low to Near-Atomic Resolution**. *Comput Struct Biotechnol J* 2015, **13**:492-503.

74. Schmidt S, Urlaub H: **Combining cryo-electron microscopy (cryo-EM) and cross-linking mass spectrometry (CX-MS) for structural elucidation of large protein assemblies**. *Curr Opin Struct Biol* 2017, **46**:157-168.

75. Rieping W, Habeck M, Nilges M: **Inferential structure determination**. *Science* 2005, **309**:303-306.

76. Habeck M: **Bayesian Modeling of Biomolecular Assemblies with Cryo-EM Maps**. *Front Mol Biosci* 2017:10.3389/fmolb.2017.00015.

77. Hummer G, Köfinger J: **Bayesian ensemble refinement by replica simulations and reweighting**. *J Chem Phys* 2016, **143**:243150.

78. Bock LV, Blau C, Vaiana AC, Grubmuller H: **Dynamic contact network between ribosomal subunits enables rapid large-scale rotation during spontaneous translocation**. *Nucl Acids Res* 2015, **43**:6747-6760.

79. Iudin A, Korir PK, Salavert-Torres J, Kleywegt GJ, Patwardhan A: **EMPIAR: a public archive for raw electron microscopy image data**. *Nat Methods* 2016, **13**:387-388.

80. Tiberti M, Papaleo E, Bengtsen T, Boomsma W, Lindorff-Larsen K: **ENCORE: Software for Quantitative Ensemble Comparison**. *PLoS Comp Biol* 2015, **11**:e1004415.

81. Burley SK, Kurisu G, Markley JL, Nakamura H, Velankar S, Berman HM, Sali A, Schwede T, Trewhella J: **PDB-Dev: A Prototype System for Depositing Integrative/Hybrid Structural Models**. *Structure* 2017, **25**:1317-1318.

82. Viswanath S, Chemmama IE, Cimermancic P, Sali A: **Assessing Exhaustiveness of Stochastic Sampling for Integrative Modeling of Macromolecular Structures**. *Biophys J* 2017, **113**:2344-2353.

83. Chen VB, Arendall III WB, Headd JJ, Keedy DA, Immormino RM, Kapral GJ, Murray LW, Richardson JS, Richardson DC: **MolProbity: all-atom structure validation for macromolecular crystallography**. *Acta Crystallogr D* 2010, **66**:12-21.

84. Adams PD, Afonine PV, Bunkoczi G, Chen VB, Echols N, Headd JJ, Hung LW, Jain S, Kapral GJ, Kunstleve RWG, et al.: **The Phenix software for automated determination of macromolecular structures**. *Methods* 2011, **55**:94-106.

85. Barad BA, Echols N, Wang RY-R, Cheng Y, DiMaio F, Adams PD, Fraser JS: **EMRinger: side chain–directed model and map validation for 3D cryo-electron microscopy**. *Nat Methods* 2015, **12**:943-946.

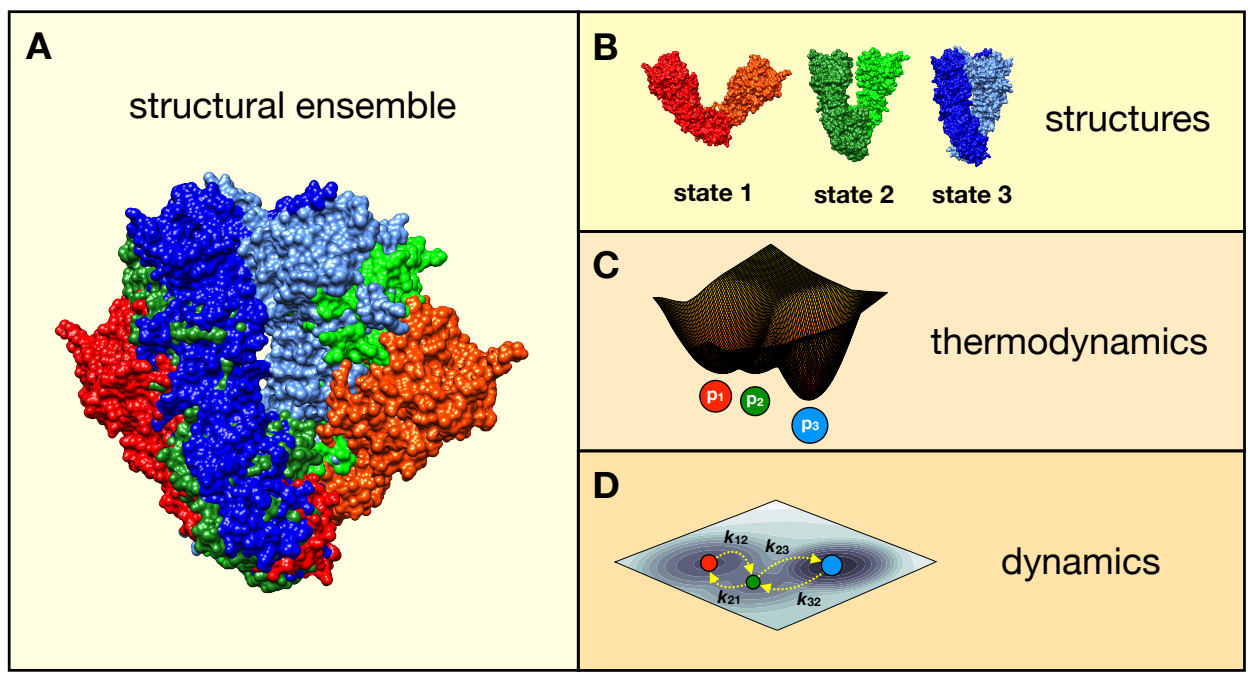

**Figure 1. Protein structural ensembles provide unique insights to understand protein behaviour.** Obtaining a comprehensive understanding of protein behaviour is facilitated by the knowledge of their structures, thermodynamics, and dynamics, which can be provided in terms of structural ensembles (A). A structural ensemble is defined by: B) the structures of all the relevant states that a protein can occupy (indicated as 1, 2, and 3) representing the open, closed, and compact conformations of the molecular chaperone Hsp90 [6]; C) their corresponding populations ($p_1$, $p_2$, and $p_3$), represented by the colored circles in the free-energy landscape; D) the rates of interconversion among states ($k_{12}$, $k_{21}$, $k_{23}$, and $k_{32}$), represented by the yellow dashed arrows in the projected free-energy landscape.

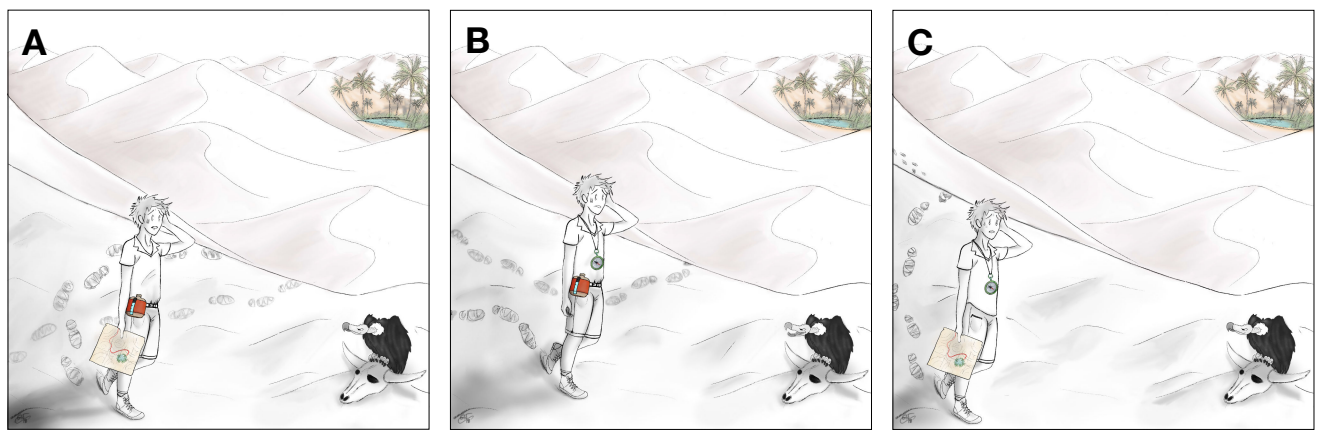

**Figure 2. Fateful challenges in protein structural ensemble determination.** Integrative methods for protein ensemble determination present three fateful challenges. A) The first challenge is obtaining an accurate energy function to drive MD or MC simulations, which should incorporate experimental and prior information on the system. This energy function could be thought as a compass, which the distressed traveller in the figure is missing, finding it essentially impossible to reach his destination. B) The second challenge consists in accurately quantifying all the sources of error and uncertainty, both in the experimental data and in the predictors used in the modelling. These are represented by the missing map in the hand of the traveller, again preventing him from arriving to the oasis. C) The final challenge is related to having enough resources to sample the conformational space of the system, or in other words achieving convergence in integrative simulations. This challenge is symbolized by the lack of water, i.e. the missing bottle at the belt of the traveller. Accurate structural ensembles, here represented by the coloured oasis, can only be obtained if the three challenges are successfully addressed. A different destiny (illustrated by the black condor) is waiting for the distressed traveller, if any of these three elements is missing. This illustration has been prepared by Giulia Vecchi.

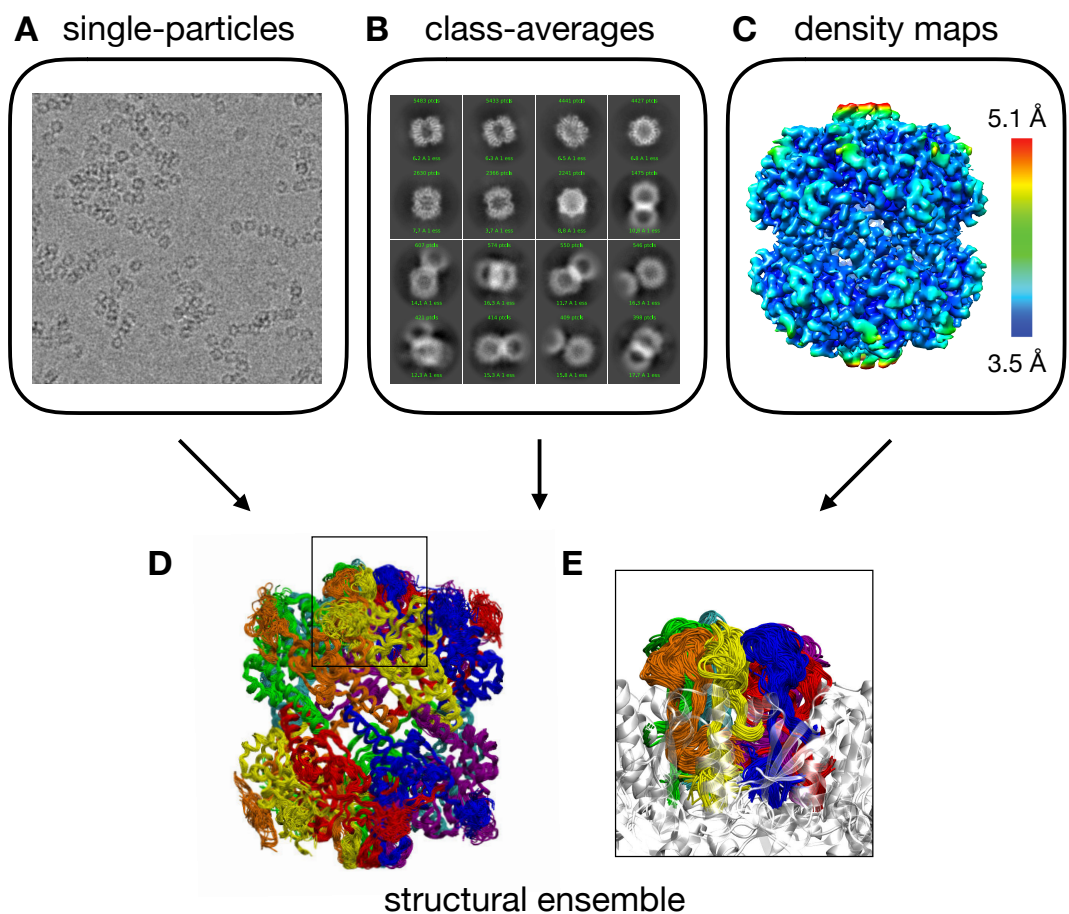

**Figure 3. Current approaches to determine protein structural ensembles using cryo-EM data.** The approaches described in this review can utilized raw, single-particle cryo-EM data (A) [55], 2D class-averages (B) [57,60], or 3D density maps [65,66] to determine structural ensembles (D), characterize the highly-dynamical regions of the system (E), and ultimately inform about protein function. This procedure is illustrated here using as an example the ClpP protease, whose cryo-EM map was determined with local resolution varying from 3.5 Å to 5.1 Å [68], as shown by the heat map in panel C. Integrative approaches can be used to combine cryo-EM with other sources of information, either using Bayesian frameworks [55,65,66] or ad-hoc hybrid energy functions [57,60].